\documentclass[10pt]{article}

\usepackage[a4paper, total={6.9in, 8.9in}]{geometry}
\usepackage{authblk}
\usepackage{amsmath,amssymb,amsfonts,bm}
\usepackage{algorithm,algorithmic}
\newcommand{\vect}[1]{\mathbf{#1}}
\newcommand{\matr}[1]{\mathbf{#1}}
\usepackage{siunitx}
\usepackage{graphicx}
\usepackage[font=footnotesize]{caption}
\usepackage{subcaption}
\usepackage{xcolor}
\usepackage{cite}

\title{Constrained B-Spline Based Everett Map Construction\\for Modeling Static Hysteresis Behavior}
\date{}

\author[1]{Bram~Daniels}
\author[2]{Timo~Overboom}
\author[1]{Reza~Zeinali}
\author[1]{Mitrofan~Curti}
\author[1]{Elena~Lomonova}
\affil[1]{Department of Electrical Engineering, Eindhoven University of Technology, 5612~AZ, Eindhoven, The~Netherlands}
\affil[2]{Royal SMIT Transformers (SGB-SMIT Group), 6531~JC, Nijmegen, The~Netherlands}

\begin{document}

\maketitle

\begin{abstract}
This work presents a simple and robust method to construct a \mbox{B-spline} based Everett map, for application in the Preisach model of hysteresis, to predict static hysteresis behavior. Its strength comes from the ability to directly capture the Everett map as a well-founded closed-form \mbox{B-spline} surface expression, while also eliminating model artifacts that plague Everett map based Preisach models. Contrary to other works, that applied numerical descriptions for the Everett map, the presented approach is of completely analytic nature. In this work the \mbox{B-spline} surface fitting procedure and the necessary set of constraints are explained. Furthermore, the \mbox{B-spline} based Everett map is validated by ensuring that model artifacts were properly eliminated. Additionally, the model was compared with four benchmark excitations. Namely, a degaussing signal, a set of first-order reversal curves, an arbitrary excitation with high-order reversal curves, and a PWM like signal. The model was able to reproduce all benchmarks with high accuracy.
\end{abstract}
\section{Introduction}
Material modeling is an integral part in the design procedure of electromechanical devices. Especially in the case of the hysteresis behavior of the ferromagnetic core, for which accurate models are desired~\cite{Ceylan2022}. The most well-known and widely adopted models to simulate the hysteresis effect of soft-magnetic materials are the Jiles-Atherton and Preisach model (PM) \cite{Jiles1986,Preisach1935}. This work opts to apply the PM, due to its capability to reproduce highly accurate model results when used in combination with a so-called Everett map~\cite{Everett1955}. The Everett map is a surface that is constructed from measured hysteresis data, \textit{e.g} concentric hysteresis loops or first-order reversal curves, which the Preisach model utilizes to form its model output~\cite{Mayergoyz2005}. However, the construction of the Everett map requires special attention to be carried out correctly, as measurement noise is easily reflected as artifacts in its output.

In the past, this was already recognized by De~Wulf et al., Novak et al., and Zeinali et al., who each construced a numerical Everett map from measurement data, and noted large errors or artifacts in the model output due to measurement noise and dissymmetry in the source data~\cite{Wulf2000,Novak2018,Zeinali2020}. These artifacts were subsequently compensated by a correction scheme or filtering. However, in a preceding work, Daniels et al. found that it is possible to eliminate most if not all measurement noise beforehand, and obtain perfectly symmetric hysteresis loops, by application of a pre-processing algorithm~\cite{Daniels2023}. Additionally, this work captured the Everett map as a \mbox{B-spline} surface, to obtain an analytic closed-form expression that allows for easy interpolation of the scattered hysteresis measurement data set. 

Nevertheless, it is not sufficient to pre-processes the measurement data, and the aforementioned correction scheme must still be applied to guarantee that the model output is free of artifacts. Especially in the case of modeled hysteresis loops that differ from the Everett map source data, \textit{e.g.} first-order reversal curves that are governed by an Everett map constructed from concentric hysteresis loops and vice versa. Evidently, for the developed hysteresis model to be generically applicable, it should be able to adequately deal with any arbitrary input. Preferably from a practical to obtain data set, \textit{i.e.} concentric hysteresis loops \textit{vs.} first-order reversal curves.

This work presents a simple and robust method to construct an Everett map from concentric hysteresis loop data, by capturing it with an analytic closed-form \mbox{B-spline} surface expression, while eliminating any artifacts in the model output. The method operates on the basis that it is possible to include the correction scheme in the fitting procedure of the \mbox{B-spline} surface, by means of a properly chosen set of constraints. This fitting procedure, as covered in a precursory work \cite{Daniels2023}, is greatly augmented by the addition of said constraints. Consequently, the constrained \mbox{B-spline} based Everett map obtained in this work, retains the highly accurate hysteresis modeling and easy interpolation of a scattered hysteresis data set, while artifact elimination is now also guaranteed, and combines this into one convenient analytic expression.

The robustness of the constructed constrained Everett map is verified by modeling a set of hysteresis loops with the PM, deliberately chosen to be very different from the concentric hysteresis loop source data from which it was constructed.

\section{PM of hysteresis}
\label{Section:model_theory}
The classical PM of hysteresis is governed by two switching variables, $\alpha$ and $\beta$, and an infinite set of hysteresis operators, $\hat{\gamma}_{\alpha,\beta}$, distributed on the triangular Preisach plane~\cite{Mayergoyz2005}, given by
\begin{equation}
f(t) = \iint \limits_{\alpha\geq\beta} \mu(\alpha,\beta) \hat{\gamma}_{\alpha,\beta} u(t) d\alpha d\beta,
\label{eq:Preisach:general}
\end{equation}
where $u(t)$ is the input signal, $f(t)$ the output signal, $\mu(\alpha,\beta)$ is the Preisach weight function or Preisach distribution, and $t$ is an instant in time. Moreover, the Everett map is incorporated by substituting the Preisach distribution and double integral for a cumulative quantity, $\xi$, as follows
\begin{equation}
\xi(u_1,u_2) = \int_{u_1}^{u_2} \int_{0}^{u_1} E(u_{1},u_{2}) du_{1} du_{2},
\label{eq:Everett_map}
\end{equation}
where $E$ is the arranged measured data, and $u_1$ and $u_2$ are the start and end point of a hysteresis branch respectively~\cite{Everett1955}. Conversely, the Preisach distribution is obtained from the Everett map by taking the double derivative of $\xi$ as follows
\begin{equation}
\mu(\alpha,\beta) = -\frac{\partial^2\xi({\alpha,\beta})}{\partial\alpha\partial\beta}.
\label{eq:Preisach_weight_function}
\end{equation}

\section{Surface representation}
This work applies a procedure similar to its precursory study to capture the Everett map as a \mbox{B-spline} surface~\cite{Daniels2023}. Hence, this section provides only the necessary descriptions, which are supplemented with several required addenda that are critical to the contributions of this work in light of constraints.

\subsection{B-spline based surface}
The non-rational \mbox{B-spline} surface is given by
\begin{equation}
\bm{S}(u,v) = \sum_{i=0}^n\sum_{j=0}^mN_{i,p}(u)N_{j,q}(v)\bm{P}_{i,j},
\label{eq:Spline_surface}
\end{equation}
where $\bm{S}(u,v)$ is the constructed surface with parametrization variables $u$ and $v$, basis functions $N_{i,p}(u)$ and $N_{j,q}(v)$ of degree $p$ and $q$ respectively, index variables $i$ and $j$ of the $n+1$ and $m+1$ basis functions, and $\bm{P}_{i,j}$ represents the control polygon~\cite{Piegl1995}. Furthermore, it is necessary to introduce the derivative of the \mbox{B-spline} basis function, to allow for constraints, which in its generic form is defined as follows
\begin{align}
N^{(k)}_{i,p}(u) = p\left(\frac{N^{(k-1)}_{i,p-1}}{u_{i+p}-u_i} - \frac{N^{(k-1)}_{i+1,p-1}}{u_{i+p+1}-u_{i+1}}\right),\label{eq:pth_basis_function}
\end{align}
where $N^{(k)}_{i,p}$ denotes the $k$th derivative of $N_{i,p}(u)$.

\subsection{Fitting procedure}
The optimal control polygon for the \mbox{B-spline} surface fit is obtained by solving a quadratic optimization problem, with a linear least-squares cost function, subjected to a set of linear constraints, as follows
\begin{equation}
\min_\vect{x} \frac{1}{2}\left\lVert\matr{C}\vect{x}-\vect{d}\right\rVert^2_2\ ~s.t.~
\begin{cases}
      \matr{A}\vect{x}\leq\vect{b},\\
      \matr{A_{eq}}\vect{x}=\vect{b_{eq}},\\
      \vect{l}\leq\vect{x}\leq\vect{u},
\end{cases} 
\label{eq:quadratic_programming}
\end{equation}
where $\matr{C}\vect{x}=\vect{d}$ is the linear system for which the norm is to be minimized, $\matr{A}\vect{x}\leq\vect{b}$ and $\matr{A_{eq}}\vect{x}=\vect{b_{eq}}$ are optional linear systems of inequality and equality constraints respectively, and $\vect{l}$ and $\vect{u}$ represent the optional lower and upper bounds\footnote{These optional bounds were not applied during fitting in this work.} for $\vect{x}$. The convex nature of the non-rational \mbox{B-spline} surface ensures reliable convergence towards the global optimal solution, \textit{i.e.} optimal control polygon\footnote{Note that for a system without constraints the solution is directly found~\cite{Daniels2023}.}.

\subsection{Constraints}
\begin{figure} 
\centering
    \begin{subfigure}{0.49\textwidth} 
		\centering
		\includegraphics[scale=1.008]{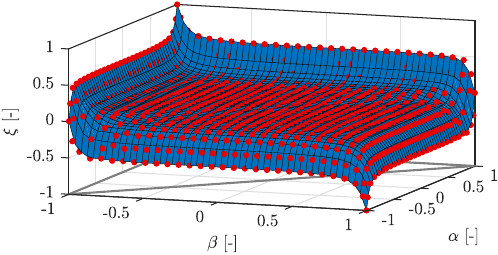}
        \caption{}
        \label{fig:spline_Everett_map}
	\end{subfigure}
	\begin{subfigure}{0.245\textwidth} 
		\centering
		\includegraphics[scale=1.008]{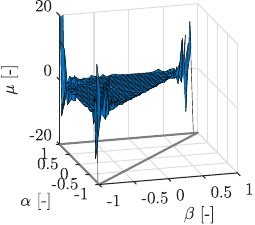}
		\caption{}
		\label{fig:negative_values_unc}
	\end{subfigure}
	\begin{subfigure}{0.245\textwidth} 
		\centering
		\includegraphics[scale=1.008]{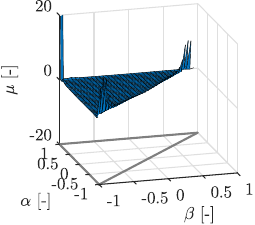}
		\caption{}
		\label{fig:negative_values_con}
	\end{subfigure}
	\caption{(a) The normalized constrained \mbox{B-spline} based Everett map, and its control points ({\color{red}$\bullet$}), with constraints applied in the Preisach plane, (b) the Preisach distribution, as determined from the unconstrained B-spline surface based Everett map, with heavy oscillations and problematic negative values, and (c) the Preisach distribution, as determined from the constrained B-spline surface based Everett map, with solely positive values.}
\end{figure}
The shape of the \mbox{B-spline} surface is manipulable during fitting by populating the inequality and equality systems of equations in \eqref{eq:quadratic_programming}. Consequently, the solver attempts to honor the applied constraints within a certain tolerance. In this way, control over the magnitude and derivative of the fitted surface, as well as the position of the control points, is obtained. For constraints on the magnitude or slope of the surface, or the position of the control points, the appropriate elements of $\matr{A_{eq}}$ are populated with the tensor basis or ones respectively, and similarly $\vect{b_{eq}}$ is populated with the desired value. For constraints on the sign of the surface slope the appropriate elements of $\matr{A}$ are populated with the derivative tensor basis
\begin{equation}
\matr{N_t^{(k,l)}}=N^{(k)}_{i,p}\otimes N^{(l)}_{j,q},
\label{eq:b_spline_tensor_basis_derivative}
\end{equation}
where $k$ and $l$ indicate the order of the derivative, and $\vect{b}$ is set to all zeros.

\section{Constrained B-spline based Everett map}
This section details about the construction of the constrained \mbox{B-spline} based Everett map. By constraining the fitting procedure it is possible to include the critical artifact elimination correction scheme into the fitting procedure, while the highly accurate hysteresis loop reproduction capabilities of the Everett map are retained. Moreover, several additional constraints are included to improve the fit. When the constrained Everett map is employed by the PM, it allows for modeling of generic hysteresis behavior from concentric hysteresis loop source data.

This work applies the following constraints:
\begin{enumerate}
  \item[1.] The Preisach distribution is non-negative,
  \item[2.] The Everett map diagonal is equal to zero,
  \item[3.] The Everett map peak is equal to magnetic saturation.
\end{enumerate}

The first constraint is an inequality constraint that eliminates the non-physical crossing of ascending and descending branches, which is a consequence of measurement noise in the Everett map source data. It is enforced on the double derivative of the fitted \mbox{B-spline} surface via \eqref{eq:Preisach_weight_function}, and ensures any negative values are removed, \textit{i.e} $\mu\geq 0$. Conveniently, it is trivial to obtain the derivative of a \mbox{B-spline} surface, which is analytically defined following \eqref{eq:pth_basis_function} and \eqref{eq:b_spline_tensor_basis_derivative}.

The second constraint is an equality constraint that removes any offset along the diagonal of the fitted \mbox{B-spline} surface, which is a consequence of the least-squares nature of the fitting procedure. It is enforced by equating the value of fitted \mbox{B-spline} surface along the diagonal of the Preisach plane to zero, and ensures a proper degaussing characteristic, which is an important property for the PM to posses.

The third constraint is an equality constraint that removes any offset at the peak value of the fitted \mbox{B-spline} surface. It is enforced by equating the value of the fitted \mbox{B-spline} surface at the peak to the value of magnetic saturation, and ensures that the PM reaches the appropriate level in magnetic saturation.

It is desirable to use a monotonically spaced knot vector for the fitted \mbox{B-spline} surface. Generally, this vector is determined by an algorithm such as the chord-length method or centripetal method~\cite{Piegl1995}. However, these often yield a non-evenly spaced knot vector, and consequently a non-linear parametrization of the \mbox{B-spline} surface, \textit{i.e.} the parametrization variables do not directly correlate to a position on the surface. Since it is critical for the PM to query exact points on the Everett map, it is preferred if the parametrization is linear. Hence, a monotonically spaced knot vector is a convenient choice\footnote{Alternatively, an implicit equation could be calculated to linearize the non-linear parametrization.}. Furthermore, it is helpful to mirror the triangular data of the Preisach plane, and obtain a square data set, that corresponds with the rectilinear basis of the \mbox{B-spline} surface. Lastly, the map was constructed with the magnetic flux density $B$ as input, and the magnetic field strength $H$ as output.

The normalized fitted constrained \mbox{B-spline} based Everett map is shown in Fig.~\ref{fig:spline_Everett_map}. The map was fitted on concentric hysteresis loop data, obtained under quasi-DC excitation over a range of \SI{0.05}{\tesla} up to \SI{1.5}{\tesla} along the rolling direction, for NO27-1450H motor steel. Furthermore, $u$ and $v$ are the parametrized variables for $\beta$ and $\alpha$ respectively. The normalization factors are $1.5$ and $1500$ for $B$ and $H$ respectively. It must be remarked that the constraints are only applied in the top half of the Everett map, \textit{i.e.} the Preisach plane  $\alpha\geq\beta$.

\section{Model validation}
\begin{figure} 
\centering
	\begin{subfigure}{0.49\textwidth} 
		\centering
		\includegraphics[scale=1]{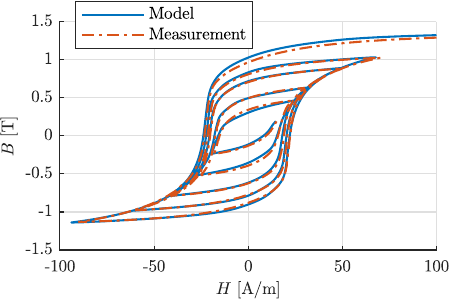}
        \caption{}
        \label{fig:model_degauss}
	\end{subfigure}
	\begin{subfigure}{0.49\textwidth} 
		\centering
		\includegraphics[scale=1]{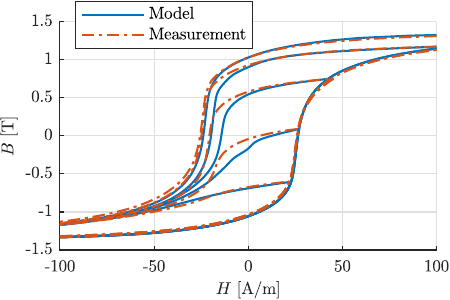}
        \caption{}
        \label{fig:model_1st_order_reversal}
	\end{subfigure}
    \caption{Model results and measurements of (a) a degaussing signal, and (b) a set of first-order reversal curves.}
\end{figure}
The model approach is validated in two ways. Firstly, to validate that the constraints were properly adhered to during the surface fitting procedure, the analytic Preisach distribution is determined by application of \eqref{eq:Preisach_weight_function} on the obtained unconstrained and constrained \mbox{B-spline} surfaces. For the unconstrained surface heavy oscillations and problematic negative values are revealed, see Fig.~\ref{fig:negative_values_unc}. Conversely, the constrained surface contains only positive values, see Fig.~\ref{fig:negative_values_con}.

\begin{figure} 
\centering
	\begin{subfigure}{0.49\textwidth} 
		\centering
		\includegraphics[scale=1]{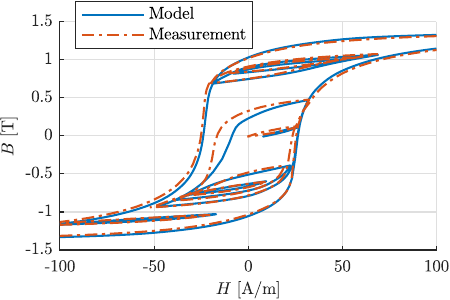}
        \caption{}
        \label{fig:hysteresis_loop}
	\end{subfigure}
	\begin{subfigure}{0.49\textwidth} 
		\centering
		\includegraphics[scale=1]{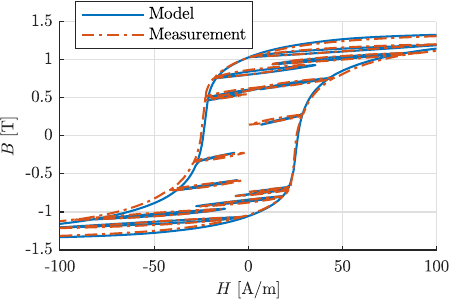}
        \caption{}
        \label{fig:model_pwm}
	\end{subfigure}
	\caption{Model results and measurements of (a) an arbitrary excitation, and (b) a PWM-like signal.}
\end{figure}

Secondly, the accuracy and robustness of the \mbox{B-spline} based PM is tested with four trial excitations, each testing a different aspect of the model. The excitations are deliberately chosen to be different from the concentric hysteresis measurements that were used to construct the Everett map, as the PM can reproduce these loops almost identically.

The first trial excitation is a degaussing signal, see Fig.~\ref{fig:model_degauss}. It is critical for a hysteresis model to properly reach the demagnetized state, as it serves as one of the three possible model start points, outside of positive and negative saturation. This is relevant when coupling with finite element software is considered. The model result exhibits no non-physical crossing of ascending and descending hysteresis branches, as was identified in~\cite{Zeinali2020}, and matches well with the measurement over the whole range of the modeled signal. However, some discrepancy is observed in the modeled $H$-field, especially for higher flux density magnitudes. Note that since a degauss signal is close in nature to concentric hysteresis loop source data, it is to be expected that this input is well reproduced.

The second trial excitation is a set of first-order reversal curves, see Fig.~\ref{fig:model_1st_order_reversal}. This excitation is of interest given that first-order reversal curves are commonly used in the identification procedure of other hysteresis models, \textit{e.g.} the congruency model, and is significantly different from the concentric hysteresis loop source data. The model result again does not exhibit any non-physical behavior, and matches well with measurements over the whole range of the modeled signal. However, some discrepancy is observed for the steeper part of the first-order reversal curves. Considering that the Everett map was created using a completely different data set, the model output still matches with the measurements remarkably well. It is expected that the model output could be further improved by incorporating more measured hysteresis loops in the surface fitting procedure, especially when additional loops of lower peak flux density values are added.

The third trial excitation is an arbitrary signal, see Fig.~\ref{fig:hysteresis_loop}. This excitation is specifically crafted such that it provides insight in the model reproduction capabilities for higher-order reversal curves. The result once more does not exhibit any non-physical behavior, and matches well with measurements over the whole range of the modeled signal. The higher-order reversal curves are reproduced to almost identical levels. However, the model produces malformed hysteresis branches around the origin, similar to the first-order reversal curves. 

The fourth and last test excitation is a PWM-like signal, which a core of an electric machine might experience, see Fig.~\ref{fig:model_pwm}. The model result again does not exhibit any non-physical behavior, and matches well with measurements over the whole range of the modeled signal. The minor loops, typical for PWM excitations, are reproduced nearly identically, even around the center area.

\section{Conclusion}
The construction of a constrained \mbox{B-spline} surface based Everett map, for usage in the PM of hysteresis, has been researched in this work. Contrary to other works, that applied numerical descriptions for the Everett map, the presented approach in this work was of completely analytic nature. The constrained fitting procedure ensured the elimination of problematic negative values in the Preisach distribution, which removed any non-physical crossing of ascending and descending hysteresis branches. The model robustness was verified by modeling four test excitations, that were deliberately chosen to be different from the concentric loop source data used in the construction of the Everett map. It has been concluded that the \mbox{B-spline} based PM was capable to accurately model all test excitations. However, a notable discrepancy was observed in the modeling of first-order reversal curves, where the model produced malformed hysteresis branches around the origin. It is likely that increasing the amount of measured loops, especially of lower peak flux density levels, can improve the model output further. The obtained constrained \mbox{B-spline} based Everett map provided highly accurate hysteresis modeling, and scattered hysteresis data interpolation capabilities, while artifacts were effectively eliminated, originating from one well-founded analytic expression.


\begin{thebibliography}{10}

\bibitem{Ceylan2022}
Doga Ceylan, Reza Zeinali, Bram Daniels, Konstantin~O. Boynov, and Elena~A.
  Lomonova.
\newblock Significance of vector hysteresis modeling in the analysis of
  variable flux reluctance machines.
\newblock In {\em 2022 International Conference on Electrical Machines
  ({ICEM})}. {IEEE}, sep 2022.

\bibitem{Daniels2023}
Bram Daniels, Timo Overboom, Mitrofan Curti, and Elena Lomonova.
\newblock Everett map construction for modeling static hysteresis:
  Delaunay-based interpolant versus b-spline surface.
\newblock {\em {IEEE} Transactions on Magnetics}, 59(5):1--4, may 2023.

\bibitem{Wulf2000}
M.~de~Wulf, L.~Vandevelde, J.~Maes, L.~Dupre, and J.~Melkebeek.
\newblock Computation of the preisach distribution function based on a measured
  everett map.
\newblock {\em IEEE Trans. Magn.}, 36(5):3141--3143, 2000.

\bibitem{Everett1955}
D.~H. Everett.
\newblock A general approach to hysteresis. part 4. an alternative formulation
  of the domain model.
\newblock {\em J. Chem. Soc. Faraday Trans.}, 51:1551, 1955.

\bibitem{Jiles1986}
D.C. Jiles and D.L. Atherton.
\newblock Theory of ferromagnetic hysteresis.
\newblock {\em J. Magn. Magn. Mater.}, 61(1-2):48--60, sep 1986.

\bibitem{Mayergoyz2005}
I.~D. Mayergoyz and Giorgio Bertotti.
\newblock {\em The Science of Hysteresis}.
\newblock Academic Press, 1 edition, 2005.

\bibitem{Novak2018}
Miroslav Novak, Jakub Eichler, and Miloslav Kosek.
\newblock Difficulty in identification of preisach hysteresis model weighting
  function using first order reversal curves method in soft magnetic materials.
\newblock {\em Applied Mathematics and Computation}, 319:469--485, feb 2018.

\bibitem{Piegl1995}
Les Piegl and Wayne Tiller.
\newblock {\em The {NURBS} Book}.
\newblock Springer Berlin Heidelberg, 1995.

\bibitem{Preisach1935}
F.~Preisach.
\newblock Über die magnetische nachwirkung.
\newblock {\em Z. Phys.}, 94(5-6):277--302, may 1935.

\bibitem{Zeinali2020}
R.~Zeinali, D.~C.~J. Krop, and E.~A. Lomonova.
\newblock Comparison of preisach and congruency-based static hysteresis models
  applied to non-oriented steels.
\newblock {\em {IEEE} Transactions on Magnetics}, 56(1):1--4, jan 2020.

\end{thebibliography}
\end{document}